\RequirePackage{ifpdf}

\documentclass[cits]{PoS}

\usepackage[authoryear]{natbib}
\bibpunct{(}{)}{;}{a}{}{,}

\title{Star and Stellar Cluster Formation: ALMA-SKA Synergies}

\ShortTitle{Star Formation}

\author{{G.~A.~Fuller}\\
  Jodrell Bank Centre for Astrophysics \& UK ALMA Regional Centre Node,
  School of Physics \& Astronomy,
  University of Manchester, UK\\
  E-mail: \email{G.Fuller@manchester.ac.uk}}
\author{J.~Forbrich\\
  Institute of Astronomy, University of Vienna, Austria\\
  E-mail: \email{jan.forbrich@univie.ac.at}}
\author{J.~M.~Rathborne\\
  CSIRO Astronomy and Space Science, Australia\\
  E-mail: \email{Jill.Rathborne@csiro.au}}
\author{S.~Longmore\\
  Astrophysics Research Institute, Liverpool John Moores University, UK\\
  E-mail: \email{S.N.Longmore@ljmu.ac.uk}}
\author{\speaker{S.~Molinari}\\
  INAF-IFSI, Rome, Italy\\
  E-mail: \email{molinari@iaps.inaf.it}}

      \abstract{{{Over the next decade, observations conducted with ALMA
and the SKA will reveal the process of mass assembly and accretion onto
young stars and will be revolutionary for studies of star formation. Here
we summarise the capabilities of ALMA and discuss recent results from its
early science observations.}}  We then review {{infrared and radio
variability observations}} of both {{young}} low-mass and high-mass stars.
A time domain SKA radio continuum survey of star forming regions is then
outlined. This survey will produce radio light-curves for hundreds of young
sources, providing for the first time a systematic survey of radio
variability across the full range of stellar masses. These light-curves
will probe the magnetospheric interactions of young binary systems, the
origins of outflows, trace episodic accretion on the central sources and
potentially constrain the rotation rates of embedded sources.}
  
\FullConference{Advancing Astrophysics with the Square Kilometre Array\\
June 8-13, 2014\\
Giardini Naxos, Italy}

\usepackage{booktabs}

\begin{document}

\makeatletter
\setbox\@firstaubox\hbox{\small Gary Fuller}
\makeatother

\section{Introduction}

The Atacama Large Millimeter/submillimeter Array
(ALMA)\footnote{www.almaobservatory.org} is the largest telescope yet
constructed.  Initial results from {{its}} early science observations
span the range from the detection of a remarkable spiral patterned
mass loss resulting { {from}} the interaction of an ABG star and its
binary companion \citep{2012Natur.490..232M} to constraining the
density of extreme starburst galaxies at $z>1$
\citep{2013MNRAS.432....2K}.  {{Here we focus on galactic star
    formation and the synergies provided by the complementarity of
    ALMA and SKA observations.}}

{ {For studies of star formation, ALMA observations will trace how
    cold dust and gas is assembled on the small-scales within a
    molecular cloud and how disks around young stars are formed.  SKA
    will be essential for complementing these studies by tracing the
    higher energy phenomena associated with the accretion of this
    material on to the star, probing the stellar and disk
    magnetospheres, and the ionising feedback from outflows and
    stellar winds. Combined, these cutting edge facilities will reveal
    the process of mass assembly and accretion onto young stars and
    will be revolutionary for studies of star formation.}}

\section{Capabilities: ALMA and SKA}

Comprising fifty four 12m diameter dishes plus twelve 7m diameter
dishes, ALMA is an international collaboration between ESO, NRAO and
NAOJ presenting Europe, North America and East Asia respectively { {in
    cooperation with the Republic of Chile}}.  Operating between
30\,GHz and 900\,GHz (10 mm -- 0.35 mm) in 10 frequency bands which
span the available atmospheric windows to study the emission from cold
dust and molecular gas, ALMA is located on the 5000m altitude
Chajnantor plateau in northern Chile.

\begin{figure}
\includegraphics{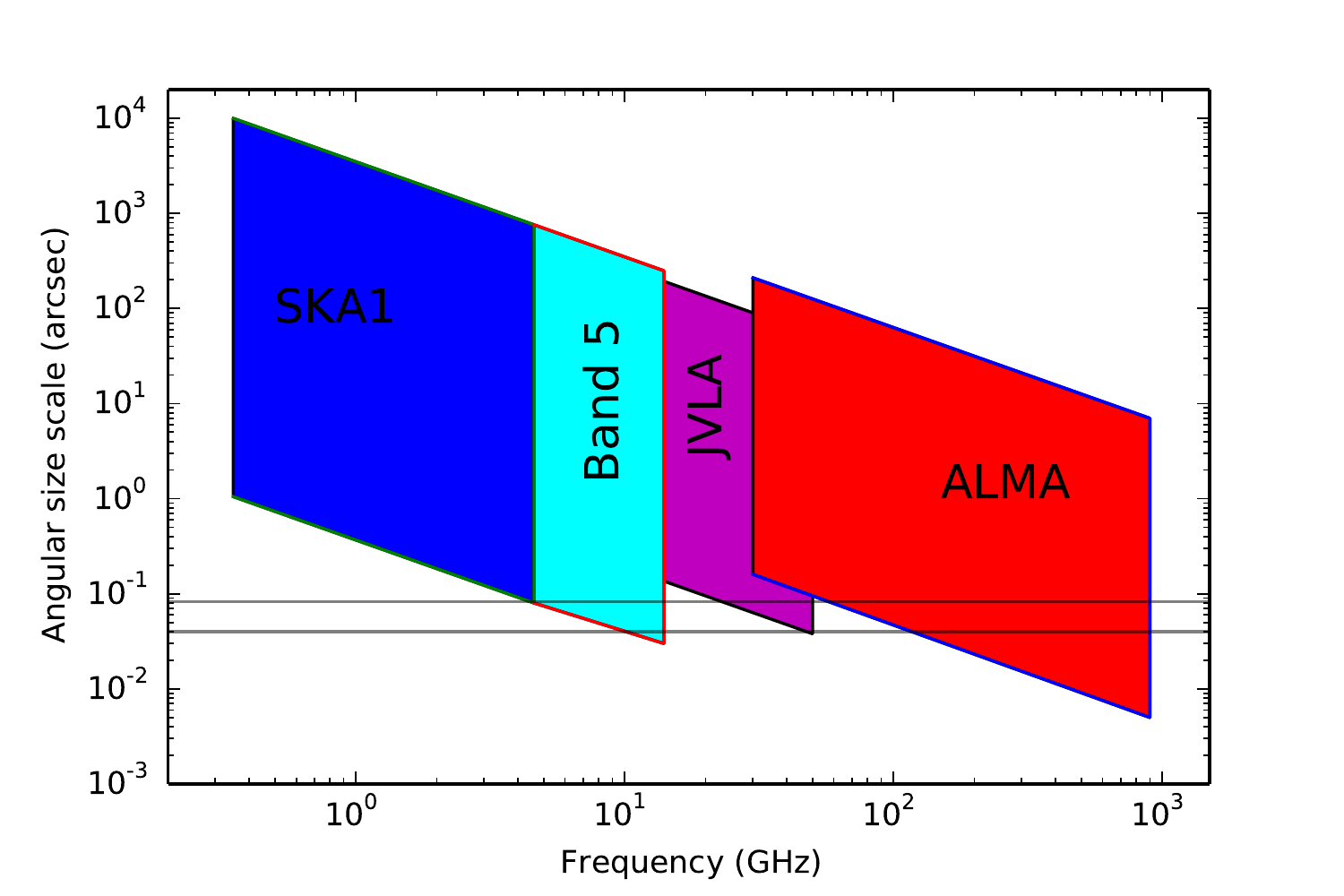}
\caption{Angular size scale as a function of frequency for SKA1, JVLA and
  ALMA. Frequency band 5 on SKA1-Mid is identified separately.  The shaded
  regions show the angular scale from the maximum resolution at each frequency
  to the instantaneous field of view. Larger angular size scales are of course
  accessible through mosaicing. The two { {horizontal}} lines indicate the
  resolution required to { {resolve}} a 0.01\,pc-sized hypercompact
  HII region (upper) and a 1000\,AU circumstellar disk/toroid {{(lower)}} at a distance of
  5\,kpc.}
\label{fig:coverage-resolution}
\end{figure}

There were two key science drivers for the design of ALMA. The first is
the imaging of the physical, chemical and magnetic structure of
protostellar/protoplanetary disks around low mass protostars in the
nearest star forming regions. {{ The second is}} imaging of Milky
Way-mass galaxies at $z=3$ in CO or C$^+$ (the brightest tracers of
molecular gas) in less than 24 hours. In parallel with these is the
requirement to routinely deliver high { {precision}} images at an
angular resolution of 0.1''. With 8\,GHz of instantaneous bandwidth,
the ALMA correlator provides spectral resolutions up to maximum of
3.8\,kHz, corresponding to a velocity resolution of 0.01\,km/s at
110\,GHz. In full operation ALMA with reach continuum noise levels of
1\,mJy (at 110\,GHz) in 1 second, corresponding to a mass of
$\sim$0.004\,M$_\odot$ per beam in the nearest star forming regions
and $\sim5$\,M$_\odot$ at 5 kpc.  Figure~\ref{fig:coverage-resolution}
shows the angular size scale versus frequency coverage of ALMA
compared with SKA1-Mid (and the NRAO Karl G. Jansky Very Large Array,
JVLA). As the figure demonstrates, to probe { {similar angular}}
scales as ALMA at frequencies above 60\,GHz, requires frequency Band 5
on SKA1-Mid.  As an indication of the range of physical scales {
  {relevant}} for studies of galactic star formation, the horizontal
lines on the figure show the angular size required to image a
hypercompact HII region, the earliest phase of massive star formation
where ionised gas is detected, and a large circumstellar disk/toroid
at the typical distance to molecular clouds in the molecular ring,
$\sim5$\,kpc.

\section{Tracing global collapse, disks and outflows}

The ultimate fate of infalling material, and how stars actually gain
mass, { {depends}} on the processes taking place {{ on small spatial
    scales close to a forming star. Recent evidence suggests that
    the}} final stages of accretion on to the central { {star may be
    highly}} episodic.  For example, ALMA observations of the chemical
composition of the envelope around a low mass protostar shows evidence
of heating associated with a burst of accretion 100-1000 years ago
when the mass accretion rate was a factor of $\sim100$ above its
current rate \citep{2013ApJ...779L..22J}.

On the large spatial scales,
ALMA observations of the massive
infrared dark cloud SDC335 \citep{2013A&A...555A.112P} show that the infalling
dense gas seen in single dish images is confined to a number of filaments, 
{{allowing us to trace the large-scale global collapse. On the small scales, 
ALMA observations of high excitation lines of both
complex organic species \citep{2012A&A...544L...7P} and simple species
\citep{2013ApJ...764L..14Z} show infall profiles towards one component of the
low mass protostellar binary IRAS16293. Indeed, measuring infall is a crucially 
important tool for understanding star formation: line profiles diagnostic of infalling gas
\citep[e.g.][]{1987A&A...186..280A} have been identified towards starless cores
\citep[e.g.][]{1999ApJ...526..788L}, low mass protostars
\citep[e.g.][]{2006ApJ...637..860W}, high mass young stellar objects
\citep[e.g.][]{2005A&A...442..949F}, UCHII regions
\citep[e.g.][]{1996ApJ...472..742H}, and molecular filaments
\citep[e.g.][]{2013ApJ...766..115K}. 
 }}

Circumstellar disks are the mass reservoir out of which planets form. ALMA
{{observations are elucidating}} the structure and composition of these disks with
unprecedented detail. Observations of transition disks, relatively evolved
disks with larger inner holes, have revealed large azimuthal asymmetries in
the dust continuum emission at radii of $\sim30-50$\,AU
\citep{2013Natur.493..191C,2013Sci...340.1199V,2014ApJ...783L..13P}.  These
asymmetries are interpreted as resulting from the trapping of dust {{ in large}}
anticyclonic vortices in the disk.  

The growth of dust {{ grains}} is a critical step towards planet
formation which SKA will probe. Regions of disks where this growth may
be occurring {{ are revealed by ALMA observations, not only of the dust
    continuum emission, but also using}} molecular lines to trace the
CO snow-line, where CO condensed on to grain surfaces
\citep{2013Sci...341..630Q, 2013A&A...557A.132M}. {{ ALMA observations
    also reveal evidence for complex chemistry in a transition disk}}
\citep{2014A&A...563A.113V} as well the presence of a simple sugars
\citep{2012ApJ...757L...4J}.  Future SKA observations will be critical
for the identification of complex organic species towards these
objects.

{{Material in protoplanetary disks}} can be eroded by
the winds and high energy radiation from nearby {{high-mass}} stars. {{To quantify this interaction, 
ALMA observations can probe molecular gas in the disk
\citep{2014ApJ...784...82M} while SKA can probe the photoionized
gas flow from the disk as well as the interaction zone between the gas and
the impinging stellar wind \citep[e.g.][]{2002ApJ...570..222G}.}}

Outflows are an ubiquitous feature of star formation.  Driven by winds and jets
which originate from the inner disk/disk-star interface region
\citep{2014prpl.conf..173L}, they not only sculpt, and eventually clear the
material around the protostar, but also inject energy into the surrounding
medium and remove angular {{momentum.}} The bulk of the
molecular line { {emission}} seen from outflows {{detected via ALMA observations}}
\citep[e.g.][]{2013ApJ...774...39A} {{ will trace material swept up by the stellar wind.}} The nature
of the jets from stars evolves as the driving source evolves
\citep{2014prpl.conf..451F} but observations show the jets have a significant
atomic component potentially observable by SKA in HI.




\section{Accretion and {{Variability in Young Stars}}}



Low-mass young stellar objects (YSOs) are thought to accrete mass via
their circumstellar disks. 
Associated accretion shocks can be observed both
when mass is accreted onto the disk and also when it is eventually accreted
onto the central object \citep[e.g.][]{2009apsf.book.....H}.  The accretion
process is observable at wavelengths ranging from the radio to the X-ray
regime, including optical and infrared wavelengths.  The connection of
accretion and high-energy processes \citep[e.g.][]{1999ARA&A..37..363F} detected
in both X-ray and radio emission has been highlighted by spectacular examples
like V1647 Ori \citep{2004Natur.430..429K}, while spectral lines in the X-ray
can provide an estimate of the the accretion rate
\citep{2012ApJ...760L..21B}.

In the centimeter { {regime}}, accretion processes can be associated {{with}} both
thermal (free-free) and non-thermal (gyrosynchrotron) radiation, as
discriminated by their spectral and polarization properties. Generally, it is
thought that the non-thermal radiation emanates from the magnetospheric
structures in the innermost vicinity of YSOs while thermal emission most
likely traces the bases of outflows and jets further away from the central
object.

Recently there has been considerable observational progress in two
relevant areas. First {{at centimeter wavelengths}}, new
instrumentation, {{provided via the}} expanded capabilities of the
JVLA, is rekindling protostellar radio astronomy. In tandem with these
advances, a multi-wavelength, time-domain view of the dynamic
processes in YSOs, including accretion is emerging.  Well sampled time
series {{datasets}} are now becoming available, in particular in the
infrared. Simultaneously obtained multi-wavelength time series
datasets are also becoming increasingly important to constrain the
underlying physical processes.

\subsection{Identifying radio counterparts to low-mass YSOs}

{{Observations with the VLA initiated}} protostellar radio astronomy
by providing excellent angular resolution to {{reliably}} identify
radio counterparts to {{low-mass}} YSOs. While the first
radio-detected YSOs identified by the VLA were found in the Orion
Nebula Cluster
\citep[e.g.][]{1983ApJ...271L..31M,1987ApJ...314..535G}, the radio
sample remained incomplete, as indicated by the fraction of radio
detected YSOs being consistently lower than the X-ray detection
fraction \citep{2013A&A...551A..56F}.

Prior to the sensitivity upgrade of the VLA, only the most nearby star-forming regions
could be studied {{in detail}}. The low-mass \textit{Coronet} cluster at a
distance of 130~pc is an example where almost all known YSOs in the inner
cluster have radio counterparts \citep{1987ApJ...322L..31B,2006A&A...446..155F}. This region was shown to host the first of only a
few known protostars with confirmed non-thermal radio emission as inferred from
circular polarization indicative of gyrosynchrotron emission
\citep{1998ApJ...494L.215F,2009ApJ...690.1901C}.

The newly upgraded JVLA has {{already significantly advanced this research by
    providing a more complete census of YSOs detected at centimetre
    wavelengths}} \citep{2013ApJ...775...63D,2014ApJ...790...49K}. {{Indeed, with the improved JVLA,}} a more
complete census of YSOs in nearby star-forming regions is emerging, both in
terms of sensitivity and in area covered.  Using rapid variability, negative
spectral indices and polarization as indicators, {{ a recent, large survey of
    $\rho$ Oph revealed that}} about half of the identified YSOs show
non-thermal emission \citep{2013ApJ...775...63D}.


Similarly, {{ a recent}} large-scale (more
than 2 square degrees) survey of the Orion Nebula Cluster, {{ detected}} 374
sources at 4.5~GHz \citep{2014ApJ...790...49K}. Of these, 148 had been previously classified as YSOs and
86 additional sources are inferred to be new YSO candidates. With reliably
determined spectral indices, these sources will be used as {{targets}} 
for follow-up VLBI observations to study parallaxes and proper motions. New
and sensitive monitoring observations of the \textit{Coronet} cluster 
{{show that the youngest}} sources in the sample are the brightest and least {{variable, possibly}}
exhibiting mostly thermal wind emission \citep{2014ApJ...780..155L}.

\subsection{Time variability}

{{The significant increase in sensitivity provided by new JVLA
    observations}} is also starting to produce radio light-curves for
YSOs. While YSO X-ray flares have been known for some time
\citep[e.g.][]{2008ApJ...688..418G}, very little is known about
protostellar radio flares
\citep{2003ApJ...598.1140B,2008A&A...477..267F} and a possible
X-ray--radio connection
\citep[e.g.][]{1993ApJ...405L..63G,2013A&A...551A..56F}. As a result
the physics of YSO radio flares is not currently well understood
although they are thought to be produced by coronal-type activity in
scenarios that also produce X-ray emission
\citep[e.g.][]{1992ApJS...82..311D}. Some of the activity could also
be directly or indirectly related to accretion.
However, some YSO flares are thought to be due to the
interaction of large magnetospheric features in close binary systems (e.g.,
V773 Tau: \cite{2006A&A...453..959M} and DQ Tau:
\cite{2008A&A...492L..21S,2011ApJ...730....6G} and references
therein) {{ or from periodic accretion bursts \citep{2013Natur.493..378M,2014ApJ...789L..38B,2014ApJ...792...64B}. }}

Impressive time domain studies are now being carried out in the infrared.
A 30-day \textit{Spitzer} and \textit{CoRot} photometric monitoring campaign
of more than a thousand YSOs in NGC~2264 {{\citep{2014AJ....147...82C}}}
classified a variety of different variability mechanisms in YSOs, mainly on
timescales of days.  {{Other recent work suggests that}} 
short-term optical variability {{may be}} due to enhanced accretion activity {{
\citep{2014AJ....147...83S}, while}} mid-infrared variability {{ may be }} due to 
structural perturbations in the inner disk {{\citep{2013AJ....145...66F}}}.
Spectroscopic monitoring campaigns have also revealed a new level of
complexity. {{Multi-epoch}} near-infrared
spectroscopy of several accreting YSOs in $\rho$ Oph {{observed}} as part of the
\textit{Spitzer} YSOVAR project, {{showed no}} correlation between the YSO mid-IR light curves and
time-resolved veiling or mass accretion rates {{\citep{2012PASP..124.1137F}}}.

\subsection{Accretion onto high-mass stars}

Strong magnetic fields and circumstellar disks are key drivers of the
evolution of, and activity in, the inner circumstellar regions of low mass
stars and there is growing evidence these important components are also
present in young high-mass, $M>8M_\odot$, stars.  Keplerian disks have been
detected through high angular resolution imaging of molecular lines around a
number of B-type stars \citep[e.g.][]{2014A&A...566A..73C,2014A&A...571A..52B}
where, like their lower mass counterparts, they presumably play a role in the
accretion of material on to the central star.  Stellar magnetic fields in the
range of hundreds to thousands of kG have also been measured towards OB stars
\citep{2014psce.conf..340N}. Since stellar magnetic fields are expected to decay
away as the high-mass star evolves due to the absence of a convective zone,
similar or stronger fields are likely present towards younger stars where they
can play an important role in the rotational braking of the stars, setting
their initial rotation rates
\citep{2011A&A...525L..11M,2012ApJ...748...97R}. The presence of disks and
strong magnetic fields suggest that young high-mass stars can show a similar
range of circumstellar energetic phenomena as seen towards young low-mass
stars, probing the final stages of mass inflow on to the central star as well
as the launching of outflows.

Indeed non-thermal radio emission has been detected towards an
increasing number of young high-mass stars
\citep[e.g.][]{1988ApJ...335..940A,2012ApJ...755..152R,2013A&A...558A.145M,2014RMxAA..50....3R}.
This emission can arise from a range of phenomena including
magnetospheric activity (such as due to variations in magnetically
mediated accretion on to the central star), interacting winds in a
massive binary system and synchrotron emission from a jet, all
processes which are likely to be variable on a range of timescales as
seen towards low mass sources \citep[e.g.][]{2013ApJ...775...63D}. For
example, a source in Orion has been seen to increase in flux by factor
of 3 increase over 1 hour \citep{2008ApJ...685..333G}.

Thermal radio emission from young high-mass stars arises from their ionised
winds and their self-photoionised disks, envelopes and surroundings in the
form of HII regions.  The accretion flow into a HII region and on to forming
massive stars will change the ionization balance and hence, flux and size of
the ionised region.  Such inflowing material will be clumpy and so the mass
accretion variable.  This is borne out by simulations which show that the
accretion rates onto stars has considerable time variability, with peak rates
between 10 and 100 times higher than the time-averaged values
\citep[e.g.][]{2012MNRAS.421.2861K}.  This variable inflow results in a flickering
of the radio continuum emission of the HII regions around massive stars. The
simulations by \cite{2011MNRAS.416.1033G} found that
about 10\% of HII regions showed flux variations of 10\% or more in 10 year
timescales. A similar result was found by \cite{2012ApJ...758..137K}. The
radio variability of HII regions can therefore probe the structure and
variability of the accretion into massive stars.

To date there have only been a very limited number of studies of the
radio variability of HII regions.  \cite{2008ApJ...674L..33G}
identified a $\sim45$\% decrease in flux (at 5\,GHz) of a hypercompact
HII (HCHII) region over a 5 year period due to enhanced accretion,
while 10\% of the ultracompact HII (UCHII) regions with Sgr B2 showed
significant changes in flux over a timescale of 23 years
\citep{2014ApJ...781L..36D}. The variable central source in W3(OH) has
changed flux by a factor of 5 on timescales of 9 years
\citep{2013ApJ...772..151D} which is interpreted as due to changes in
the ionised atmosphere of a circumstellar disk possibly due to changes
in the accretion through the circumstellar disk.

The masers observed towards many embedded massive stars provide an additional
sensitive probe of changes in the continuum emission of their exciting
sources. For example, towards some sources the 6.7\,GHz methanol maser
emission varies periodically or quasi-periodically
\citep{2014MNRAS.437.1808G,2014MNRAS.439..407S} which may reflect variations
in either the free-free background radiation due to changes in the ionization
of the circumstellar material or changes in the infrared pump radiation. Both
of these effects may arise from periodic accretion from a circumstellar disk.

\section{SKA1 Young Star Variability Survey}

Both low- and high-mass young stars show centimetre radio continuum
emission which is variable on a range of timescales from sub-hour to
years and decades.  This emission traces a range of phenomena
associated with accretion, magnetospheric activity, and outflow which
link the hot inner circumstellar regions with the infall and outflow
of cooler material traced on larger scales by ALMA.  SKA time domain
continuum surveys of young stars will provide powerful probes of the
inner circumstellar regions of young stars, opening new windows on
the star formation process.  For example, providing the first
comprehensive studies of the properties of the episodic accretion in
the inner circumstellar regions of forming massive stars.

\begin{figure}
\includegraphics{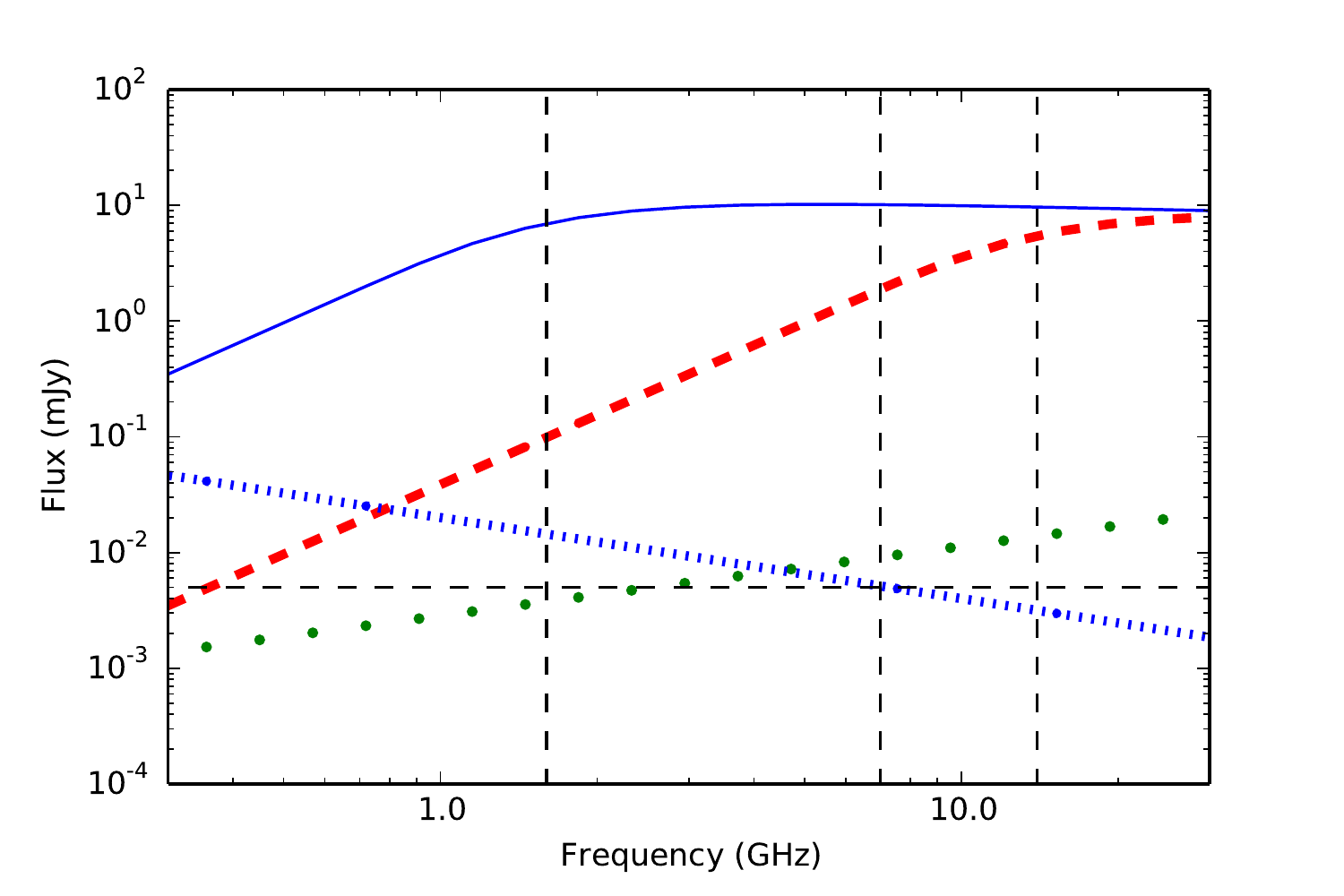}
\caption{{{Spectral flux distribution for typical }} UCHII region (solid, blue curve) with an
  emission measure of $10^7$\,pc\,cm$^{-6}$, a typical HCHII region (dashed,
  red curve) with an emission measure of $10^9$ \,pc\,cm$^{-6}$, a typical
  ionised wind (with the flux scaling as $\nu^{0.6}$) (green dots) and
  synchrotron emission $\propto \nu^{-0.7}$ (dotted, blue curve).  The break
  in the spectrum of the UCHII and HCHII regions marks the transition between
  optically thin and optically thick emission.  The wind and synchrotron
  emission are normalised to the 4.5 GHz measured flux of young stellar
  objects detected in Orion \citep{2014ApJ...790...49K} scaled to a distance
  of 5 kpc. The vertical dashed lines indicate the survey frequencies while
  the horizontal dashed lines shows the survey 5-$\sigma$ noise level
  (5\,$\mu$Jy).}
\label{fig:spec}
\end{figure}

In order to confirm the association of radio emission with a particular YSO in
{{crowded cluster environments}}, as well as isolate the emission {{close to
    the central star}}, such a SKA survey will require sub-arcsecond angular
resolution. In addition it will require near simultaneous wide multi-frequency
coverage (Figure~\ref{fig:spec}) to constrain the spectral index of {{the
    emission to discriminate}} between thermal and non-thermal emission as
well as optically thick and optically thin emission. Measurements of the
thermal emission in both the optically thick and thin regimes are important to
distinguish between changes in size of the HII region and its ionization. Full
Stokes synthesis will be important not only for confirmation of the presence
of non-thermal emission, but also to allow the separation of thermal and
non-thermal components of the emission.

Figure~\ref{fig:spec} {{shows the spectral flux distribution}} of
young HII regions as well as thermal wind and synchrotron emission
from lower mass stars.  To distinguish between the possible emission
mechanisms, observations at three frequencies are required.  A
proposed set of survey frequencies, 1.6\,GHz, 7\,GHz and 14\,GHz, is
shown on the figure. As well as providing good sampling of the
continuum spectrum of the sources, these particular frequencies also
cover several of the key maser transitions (ground state OH masers at
1.6\,GHz, Class II methanol masers at 6.7\,GHz and 12.2\,GHz, and the
H$_2$CO 2(1,1)-- 2(1, 2) transition at 14.5\,GHz\footnote{This line is
  not currently expected to be covered by SKA1 although this would be 
  desirable. However, this frequency is expected to be available with
  the full SKA.})  which are themselves variable, in some cases
periodically, and provide an additional, sensitive tracer of changes
in the emission from the central sources.  A summary of the parameters
and the required observing time for a sample survey are given in
Table~\ref{tab:survey}.

\begin{table}[h]
\begin{tabular}{cccccp{7cm}}
\toprule
Frequency & F$_{\rm src}$ & $\sigma$ & $\theta$ & Time & Comments \\
(GHz) & ($\mu$Jy) & ($\mu$Jy) & ('') & (hours/field) & \\
\midrule
1.6 & 14 & 1 & 0.5 &5 & Optically thick UCHII regions; synchrotron; OH ground state masers\\
7 & 9 & 1 & 0.5 &2.5 & Optically thin UCHII, thick HCHII; synchrotron-thermal wind cross over; 6.7\,GHz methanol masers\\
14 & 14 & 1 & 0.5 & 2.5& Low optical depth HCHII; thermal wind emission;
12.2\,GHz methanol masers; H$_2$CO transition\\
\bottomrule
\end{tabular}
\caption{SKA1 young star variability survey parameters. The columns
  show the frequency of observation, the target source flux (F$_{\rm
    src}$) and rms flux noise levels ($\sigma$), the required angular
  resolution ($\theta$) and the integration time per field. The comments indidate the emission probed and spectral lines which would be covered in the observations. }
\label{tab:survey}
\end{table}

To sample the range of physical phenomena which give rise to variable
radio emission requires repeated observations with a wide range of
cadences. Analysis of subsets of the observations will sample sub-hour
variability due to flares arising from accretion or magnetospheric
events \citep[e.g.][]{2008A&A...492L..21S,2006A&A...453..959M} and
rapidly rotating stars \citep{2006AJ....132..749W} while daily to
monthly observations will sample rotating stars and binary systems
with longer periods. Longer observation intervals will sample
processes ranging from outflow bursts to changes in the ionization,
and therefore flickering, of HII regions due to accretion flows
\citep{2011MNRAS.416.1033G}. 

The angular resolution of the survey is set by the requirement to isolate
individual sources in crowded star forming regions and resolve the emission
from individual HCHII regions with sizes $\sim0.02$\,pc out to at least the
molecular ring at $\sim5$\,kpc.  To connect the detailed studies which can be
carried out in nearby star forming regions with the processes in the wider
galactic plane requires that similar sources are detectable in both kinds of
regions. Therefore the sensitivity required for each single observation is set
by the need to sample a similar range of sources to those observed in nearby
star forming regions such as Orion at the typical distance of a star forming
region in the molecular ring, 5\,kpc. To detect wind emission sources and
synchrotron sources in the Orion star forming region
\citep{2014ApJ...790...49K} at 5\,kpc implies a required noise level of
1\,$\mu$Jy at each of the observing frequencies (Fig. 2) which should provide
detections of up to few hundred sources per star forming region.

Combining the observations at the different epochs will also provide a
highly sensitive survey for faint sources including HII regions around
relatively low mass stars. The combination of 14\,GHz observations at
10 epochs would reach a noise level of $\sim0.3$\,$\mu$Jy, sensitive
at a 6-$\sigma$ level to the HII regions from stars with  masses
as low as 6.5M$_\odot$ corresponding to a spectral type of B5
\citep{1998ApJ...501..192D}.

\section{Summary}

A SKA time-domain survey of star forming regions will produce well
sampled radio light-curves of hundreds of sources per star forming
region.  As is the case for infrared light-curves of young sources,
these will undoubtably have a range of properties, tracing and
constraining a range of phenomena. For example, the identification of
rotationally modulated emission will provide the first observational
constraints on the rotation rates of embedded protostars. Observations
of the flickering of HII regions will trace the inflow of material
around the highest mass objects while flare activity in these, and
other, sources can trace the final infall of material on to the
central protostars. These observations can not only constrain the
properties of the accretion but also the stochastic heating of the
circumstellar regions which can surpress fragmentation
\citep[e.g.][]{2007ApJ...656..959K}, leading to the formation of more
massive stars.

Looking forward to full SKA the factor 10 enhancement in sensitivity
will enable studies of smaller (and therefore also shorter time scale) flux
variations. Full SKA will also make stellar clusters to beyond the Galactic centre
accessible to detailed time-domain studies, probing star formation over a much
wider range of environments. 
The enhanced
resolution of SKA will allow studies of the changes in morphology of the
sources during their variation in flux, provide stronger constraints on the
mechanisms involved and the models for the emission.

Building a comprehensive model for the formation of both low-mass and
high-mass stars requires understanding the evolution of gas and dust
from molecular clouds down through clumps and cores and eventually on
to the forming stars.  This is only possible with the combination of
ALMA to trace the cool molecular gas and dust and SKA to follow this
material into the inner circumstellar regions. The radio light-curves
which SKA will produce for hundreds of sources in star forming regions
out to 5 kpc will provide the first detailed insight across a wide
range of stellar masses of the transient energetic phenomena occurring
on the small spatial scales close to the central star. These light
curves will for the first time provide a systematic survey which can
study the magnetospheric interactions in young binary systems, tracing
episodic accretion, the origin of outflows and potentially constrain
the rotation rates of deeply embedded sources.

\setlength{\bibsep}{0.0pt}
\bibliography{ska}{}

\begin{thebibliography}{}
\expandafter\ifx\csname natexlab\endcsname\relax\def\natexlab#1{#1}\fi

\bibitem[{{Andre} {et~al.}(1988){Andre}, {Montmerle}, {Feigelson}, {Stine}, \&
  {Klein}}]{1988ApJ...335..940A}
{Andre}, P., {Montmerle}, T., {Feigelson}, E.~D., {Stine}, P.~C., \& {Klein},
  K.-L. 1988, \apj, 335, 940

\bibitem[{{Anglada} {et~al.}(1987){Anglada}, {Rodriguez}, {Canto}, {Estalella},
  \& {Lopez}}]{1987A&A...186..280A}
{Anglada}, G., {Rodriguez}, L.~F., {Canto}, J., {Estalella}, R., \& {Lopez}, R.
  1987, \aap, 186, 280

\bibitem[{{Arce} {et~al.}(2013){Arce}, {Mardones}, {Corder}, {Garay},
  {Noriega-Crespo}, \& {Raga}}]{2013ApJ...774...39A}
{Arce}, H.~G., {Mardones}, D., {Corder}, S.~A., {et~al.} 2013, \apj, 774, 39

\bibitem[{{Balog} {et~al.}(2014){Balog}, {Muzerolle}, {Flaherty}, {Detre},
  {Bouwmann}, {Furlan}, {Gutermuth}, {Juhasz}, {Bally}, {Nielbock}, {Klaas},
  {Krause}, {Henning}, \& {Marton}}]{2014ApJ...789L..38B}
{Balog}, Z., {Muzerolle}, J., {Flaherty}, K., {et~al.} 2014, \apjl, 789, L38

\bibitem[{{Bary} \& {Petersen}(2014)}]{2014ApJ...792...64B}
{Bary}, J.~S., \& {Petersen}, M.~S. 2014, \apj, 792, 64

\bibitem[{{Beltr{\'a}n} {et~al.}(2014){Beltr{\'a}n}, {S{\'a}nchez-Monge},
  {Cesaroni}, {Kumar}, {Galli}, {Walmsley}, {Etoka}, {Furuya}, {Moscadelli},
  {Stanke}, {van der Tak}, {Vig}, {Wang}, {Zinnecker}, {Elia}, \&
  {Schisano}}]{2014A&A...571A..52B}
{Beltr{\'a}n}, M.~T., {S{\'a}nchez-Monge}, {\'A}., {Cesaroni}, R., {et~al.}
  2014, \aap, 571, A52

\bibitem[{{Bower} {et~al.}(2003){Bower}, {Plambeck}, {Bolatto}, {McCrady},
  {Graham}, {de Pater}, {Liu}, \& {Baganoff}}]{2003ApJ...598.1140B}
{Bower}, G.~C., {Plambeck}, R.~L., {Bolatto}, A., {et~al.} 2003, \apj, 598,
  1140

\bibitem[{{Brickhouse} {et~al.}(2012){Brickhouse}, {Cranmer}, {Dupree},
  {G{\"u}nther}, {Luna}, \& {Wolk}}]{2012ApJ...760L..21B}
{Brickhouse}, N.~S., {Cranmer}, S.~R., {Dupree}, A.~K., {et~al.} 2012, \apjl,
  760, L21

\bibitem[{{Brown}(1987)}]{1987ApJ...322L..31B}
{Brown}, A. 1987, \apjl, 322, L31

\bibitem[{{Casassus} {et~al.}(2013){Casassus}, {van der Plas}, {M}, {Dent},
  {Fomalont}, {Hagelberg}, {Hales}, {Jord{\'a}n}, {Mawet}, {M{\'e}nard},
  {Wootten}, {Wilner}, {Hughes}, {Schreiber}, {Girard}, {Ercolano}, {Canovas},
  {Rom{\'a}n}, \& {Salinas}}]{2013Natur.493..191C}
{Casassus}, S., {van der Plas}, G., {M}, S.~P., {et~al.} 2013, \nat, 493, 191

\bibitem[{{Cesaroni} {et~al.}(2014){Cesaroni}, {Galli}, {Neri}, \&
  {Walmsley}}]{2014A&A...566A..73C}
{Cesaroni}, R., {Galli}, D., {Neri}, R., \& {Walmsley}, C.~M. 2014, \aap, 566,
  A73

\bibitem[{{Choi} {et~al.}(2009){Choi}, {Tatematsu}, {Hamaguchi}, \&
  {Lee}}]{2009ApJ...690.1901C}
{Choi}, M., {Tatematsu}, K., {Hamaguchi}, K., \& {Lee}, J.-E. 2009, \apj, 690,
  1901

\bibitem[{{Cody} {et~al.}(2014){Cody}, {Stauffer}, {Baglin}, {Micela},
  {Rebull}, {Flaccomio}, {Morales-Calder{\'o}n}, {Aigrain}, {Bouvier},
  {Hillenbrand}, {Gutermuth}, {Song}, {Turner}, {Alencar}, {Zwintz},
  {Plavchan}, {Carpenter}, {Findeisen}, {Carey}, {Terebey}, {Hartmann},
  {Calvet}, {Teixeira}, {Vrba}, {Wolk}, {Covey}, {Poppenhaeger}, {G{\"u}nther},
  {Forbrich}, {Whitney}, {Affer}, {Herbst}, {Hora}, {Barrado}, {Holtzman},
  {Marchis}, {Wood}, {Medeiros Guimar{\~a}es}, {Lillo Box}, {Gillen},
  {McQuillan}, {Espaillat}, {Allen}, {D'Alessio}, \&
  {Favata}}]{2014AJ....147...82C}
{Cody}, A.~M., {Stauffer}, J., {Baglin}, A., {et~al.} 2014, \aj, 147, 82

\bibitem[{{De Pree} {et~al.}(2014){De Pree}, {Peters}, {Mac Low}, {Wilner},
  {Goss}, {Galv{\'a}n-Madrid}, {Keto}, {Klessen}, \&
  {Monsrud}}]{2014ApJ...781L..36D}
{De Pree}, C.~G., {Peters}, T., {Mac Low}, M.-M., {et~al.} 2014, \apjl, 781,
  L36

\bibitem[{{Diaz-Miller} {et~al.}(1998){Diaz-Miller}, {Franco}, \&
  {Shore}}]{1998ApJ...501..192D}
{Diaz-Miller}, R.~I., {Franco}, J., \& {Shore}, S.~N. 1998, \apj, 501, 192

\bibitem[{{Drake} {et~al.}(1992){Drake}, {Simon}, \&
  {Linsky}}]{1992ApJS...82..311D}
{Drake}, S.~A., {Simon}, T., \& {Linsky}, J.~L. 1992, \apjs, 82, 311

\bibitem[{{Dzib} {et~al.}(2013{\natexlab{a}}){Dzib}, {Rodr{\'{\i}}guez-Garza},
  {Rodr{\'{\i}}guez}, {Kurtz}, {Loinard}, {Zapata}, \&
  {Lizano}}]{2013ApJ...772..151D}
{Dzib}, S.~A., {Rodr{\'{\i}}guez-Garza}, C.~B., {Rodr{\'{\i}}guez}, L.~F.,
  {et~al.} 2013{\natexlab{a}}, \apj, 772, 151

\bibitem[{{Dzib} {et~al.}(2013{\natexlab{b}}){Dzib}, {Loinard}, {Mioduszewski},
  {Rodr{\'{\i}}guez}, {Ortiz-Le{\'o}n}, {Pech}, {Rivera}, {Torres}, {Boden},
  {Hartmann}, {Evans}, {Brice{\~n}o}, \& {Tobin}}]{2013ApJ...775...63D}
{Dzib}, S.~A., {Loinard}, L., {Mioduszewski}, A.~J., {et~al.}
  2013{\natexlab{b}}, \apj, 775, 63

\bibitem[{{Faesi} {et~al.}(2012){Faesi}, {Covey}, {Gutermuth},
  {Morales-Calder{\'o}n}, {Stauffer}, {Plavchan}, {Rebull}, {Song}, \&
  {Lloyd}}]{2012PASP..124.1137F}
{Faesi}, C.~M., {Covey}, K.~R., {Gutermuth}, R., {et~al.} 2012, \pasp, 124,
  1137

\bibitem[{{Feigelson} {et~al.}(1998){Feigelson}, {Carkner}, \&
  {Wilking}}]{1998ApJ...494L.215F}
{Feigelson}, E.~D., {Carkner}, L., \& {Wilking}, B.~A. 1998, \apjl, 494, L215

\bibitem[{{Feigelson} \& {Montmerle}(1999)}]{1999ARA&A..37..363F}
{Feigelson}, E.~D., \& {Montmerle}, T. 1999, \araa, 37, 363

\bibitem[{{Flaherty} {et~al.}(2013){Flaherty}, {Muzerolle}, {Rieke},
  {Gutermuth}, {Balog}, {Herbst}, \& {Megeath}}]{2013AJ....145...66F}
{Flaherty}, K.~M., {Muzerolle}, J., {Rieke}, G., {et~al.} 2013, \aj, 145, 66

\bibitem[{{Forbrich} {et~al.}(2008){Forbrich}, {Menten}, \&
  {Reid}}]{2008A&A...477..267F}
{Forbrich}, J., {Menten}, K.~M., \& {Reid}, M.~J. 2008, \aap, 477, 267

\bibitem[{{Forbrich} {et~al.}(2006){Forbrich}, {Preibisch}, \&
  {Menten}}]{2006A&A...446..155F}
{Forbrich}, J., {Preibisch}, T., \& {Menten}, K.~M. 2006, \aap, 446, 155

\bibitem[{{Forbrich} \& {Wolk}(2013)}]{2013A&A...551A..56F}
{Forbrich}, J., \& {Wolk}, S.~J. 2013, \aap, 551, A56

\bibitem[{{Frank} {et~al.}(2014){Frank}, {Ray}, {Cabrit}, {Hartigan}, {Arce},
  {Bacciotti}, {Bally}, {Benisty}, {Eisl{\"o}ffel}, {G{\"u}del}, {Lebedev},
  {Nisini}, \& {Raga}}]{2014prpl.conf..451F}
{Frank}, A., {Ray}, T.~P., {Cabrit}, S., {et~al.} 2014, Protostars and Planets
  VI, 451

\bibitem[{{Fuller} {et~al.}(2005){Fuller}, {Williams}, \&
  {Sridharan}}]{2005A&A...442..949F}
{Fuller}, G.~A., {Williams}, S.~J., \& {Sridharan}, T.~K. 2005, \aap, 442, 949

\bibitem[{{Galv{\'a}n-Madrid} {et~al.}(2011){Galv{\'a}n-Madrid}, {Peters},
  {Keto}, {Mac Low}, {Banerjee}, \& {Klessen}}]{2011MNRAS.416.1033G}
{Galv{\'a}n-Madrid}, R., {Peters}, T., {Keto}, E.~R., {et~al.} 2011, \mnras,
  416, 1033

\bibitem[{{Galv{\'a}n-Madrid} {et~al.}(2008){Galv{\'a}n-Madrid},
  {Rodr{\'{\i}}guez}, {Ho}, \& {Keto}}]{2008ApJ...674L..33G}
{Galv{\'a}n-Madrid}, R., {Rodr{\'{\i}}guez}, L.~F., {Ho}, P.~T.~P., \& {Keto},
  E. 2008, \apjl, 674, L33

\bibitem[{{Garay} {et~al.}(1987){Garay}, {Moran}, \&
  {Reid}}]{1987ApJ...314..535G}
{Garay}, G., {Moran}, J.~M., \& {Reid}, M.~J. 1987, \apj, 314, 535

\bibitem[{{Getman} {et~al.}(2011){Getman}, {Broos}, {Salter}, {Garmire}, \&
  {Hogerheijde}}]{2011ApJ...730....6G}
{Getman}, K.~V., {Broos}, P.~S., {Salter}, D.~M., {Garmire}, G.~P., \&
  {Hogerheijde}, M.~R. 2011, \apj, 730, 6

\bibitem[{{Getman} {et~al.}(2008){Getman}, {Feigelson}, {Broos}, {Micela}, \&
  {Garmire}}]{2008ApJ...688..418G}
{Getman}, K.~V., {Feigelson}, E.~D., {Broos}, P.~S., {Micela}, G., \&
  {Garmire}, G.~P. 2008, \apj, 688, 418

\bibitem[{{Goedhart} {et~al.}(2014){Goedhart}, {Maswanganye}, {Gaylard}, \&
  {van der Walt}}]{2014MNRAS.437.1808G}
{Goedhart}, S., {Maswanganye}, J.~P., {Gaylard}, M.~J., \& {van der Walt},
  D.~J. 2014, \mnras, 437, 1808

\bibitem[{{G{\'o}mez} {et~al.}(2008){G{\'o}mez}, {Rodr{\'{\i}}guez}, {Loinard},
  {Lizano}, {Allen}, {Poveda}, \& {Menten}}]{2008ApJ...685..333G}
{G{\'o}mez}, L., {Rodr{\'{\i}}guez}, L.~F., {Loinard}, L., {et~al.} 2008, \apj,
  685, 333

\bibitem[{{Graham} {et~al.}(2002){Graham}, {Meaburn}, {Garrington}, {O'Brien},
  {Henney}, \& {O'Dell}}]{2002ApJ...570..222G}
{Graham}, M.~F., {Meaburn}, J., {Garrington}, S.~T., {et~al.} 2002, \apj, 570,
  222

\bibitem[{{Guedel} \& {Benz}(1993)}]{1993ApJ...405L..63G}
{Guedel}, M., \& {Benz}, A.~O. 1993, \apjl, 405, L63

\bibitem[{{Hartmann}(2009)}]{2009apsf.book.....H}
{Hartmann}, L. 2009, {Accretion Processes in Star Formation: Second Edition}
  (Cambridge University Press)

\bibitem[{{Ho} \& {Young}(1996)}]{1996ApJ...472..742H}
{Ho}, P.~T.~P., \& {Young}, L.~M. 1996, \apj, 472, 742

\bibitem[{{J{\o}rgensen} {et~al.}(2012){J{\o}rgensen}, {Favre}, {Bisschop},
  {Bourke}, {van Dishoeck}, \& {Schmalzl}}]{2012ApJ...757L...4J}
{J{\o}rgensen}, J.~K., {Favre}, C., {Bisschop}, S.~E., {et~al.} 2012, \apjl,
  757, L4

\bibitem[{{J{\o}rgensen} {et~al.}(2013){J{\o}rgensen}, {Visser}, {Sakai},
  {Bergin}, {Brinch}, {Harsono}, {Lindberg}, {van Dishoeck}, {Yamamoto},
  {Bisschop}, \& {Persson}}]{2013ApJ...779L..22J}
{J{\o}rgensen}, J.~K., {Visser}, R., {Sakai}, N., {et~al.} 2013, \apjl, 779,
  L22

\bibitem[{{Karim} {et~al.}(2013){Karim}, {Swinbank}, {Hodge}, {Smail},
  {Walter}, {Biggs}, {Simpson}, {Danielson}, {Alexander}, {Bertoldi}, {de
  Breuck}, {Chapman}, {Coppin}, {Dannerbauer}, {Edge}, {Greve}, {Ivison},
  {Knudsen}, {Menten}, {Schinnerer}, {Wardlow}, {Wei{\ss}}, \& {van der
  Werf}}]{2013MNRAS.432....2K}
{Karim}, A., {Swinbank}, A.~M., {Hodge}, J.~A., {et~al.} 2013, \mnras, 432, 2

\bibitem[{{Kastner} {et~al.}(2004){Kastner}, {Richmond}, {Grosso}, {Weintraub},
  {Simon}, {Frank}, {Hamaguchi}, {Ozawa}, \& {Henden}}]{2004Natur.430..429K}
{Kastner}, J.~H., {Richmond}, M., {Grosso}, N., {et~al.} 2004, \nat, 430, 429

\bibitem[{{Kirk} {et~al.}(2013){Kirk}, {Myers}, {Bourke}, {Gutermuth},
  {Hedden}, \& {Wilson}}]{2013ApJ...766..115K}
{Kirk}, H., {Myers}, P.~C., {Bourke}, T.~L., {et~al.} 2013, \apj, 766, 115

\bibitem[{{Klassen} {et~al.}(2012{\natexlab{a}}){Klassen}, {Peters}, \&
  {Pudritz}}]{2012ApJ...758..137K}
{Klassen}, M., {Peters}, T., \& {Pudritz}, R.~E. 2012{\natexlab{a}}, \apj, 758,
  137

\bibitem[{{Klassen} {et~al.}(2012{\natexlab{b}}){Klassen}, {Pudritz}, \&
  {Peters}}]{2012MNRAS.421.2861K}
{Klassen}, M., {Pudritz}, R.~E., \& {Peters}, T. 2012{\natexlab{b}}, \mnras,
  421, 2861

\bibitem[{{Kounkel} {et~al.}(2014){Kounkel}, {Hartmann}, {Loinard},
  {Mioduszewski}, {Dzib}, {Ortiz-Le{\'o}n}, {Rodr{\'{\i}}guez}, {Pech},
  {Rivera}, {Torres}, {Boden}, {Evans}, {Brice{\~n}o}, \&
  {Tobin}}]{2014ApJ...790...49K}
{Kounkel}, M., {Hartmann}, L., {Loinard}, L., {et~al.} 2014, \apj, 790, 49

\bibitem[{{Krumholz} {et~al.}(2007){Krumholz}, {Klein}, \&
  {McKee}}]{2007ApJ...656..959K}
{Krumholz}, M.~R., {Klein}, R.~I., \& {McKee}, C.~F. 2007, \apj, 656, 959

\bibitem[{{Lee} {et~al.}(1999){Lee}, {Myers}, \&
  {Tafalla}}]{1999ApJ...526..788L}
{Lee}, C.~W., {Myers}, P.~C., \& {Tafalla}, M. 1999, \apj, 526, 788

\bibitem[{{Li} {et~al.}(2014){Li}, {Banerjee}, {Pudritz}, {J{\o}rgensen},
  {Shang}, {Krasnopolsky}, \& {Maury}}]{2014prpl.conf..173L}
{Li}, Z.-Y., {Banerjee}, R., {Pudritz}, R.~E., {et~al.} 2014, Protostars and
  Planets VI, 173

\bibitem[{{Liu} {et~al.}(2014){Liu}, {Galv{\'a}n-Madrid}, {Forbrich},
  {Rodr{\'{\i}}guez}, {Takami}, {Costigan}, {Manara}, {Yan}, {Karr}, {Chou},
  {Ho}, \& {Zhang}}]{2014ApJ...780..155L}
{Liu}, H.~B., {Galv{\'a}n-Madrid}, R., {Forbrich}, J., {et~al.} 2014, \apj,
  780, 155

\bibitem[{{Maercker} {et~al.}(2012){Maercker}, {Mohamed}, {Vlemmings},
  {Ramstedt}, {Groenewegen}, {Humphreys}, {Kerschbaum}, {Lindqvist},
  {Olofsson}, {Paladini}, {Wittkowski}, {de Gregorio-Monsalvo}, \&
  {Nyman}}]{2012Natur.490..232M}
{Maercker}, M., {Mohamed}, S., {Vlemmings}, W.~H.~T., {et~al.} 2012, \nat, 490,
  232

\bibitem[{{Mann} {et~al.}(2014){Mann}, {Di Francesco}, {Johnstone}, {Andrews},
  {Williams}, {Bally}, {Ricci}, {Hughes}, \& {Matthews}}]{2014ApJ...784...82M}
{Mann}, R.~K., {Di Francesco}, J., {Johnstone}, D., {et~al.} 2014, \apj, 784,
  82

\bibitem[{{Massi} {et~al.}(2006){Massi}, {Forbrich}, {Menten},
  {Torricelli-Ciamponi}, {Neidh{\"o}fer}, {Leurini}, \&
  {Bertoldi}}]{2006A&A...453..959M}
{Massi}, M., {Forbrich}, J., {Menten}, K.~M., {et~al.} 2006, \aap, 453, 959

\bibitem[{{Mathews} {et~al.}(2013){Mathews}, {Klaassen}, {Juh{\'a}sz},
  {Harsono}, {Chapillon}, {van Dishoeck}, {Espada}, {de Gregorio-Monsalvo},
  {Hales}, {Hogerheijde}, {Mottram}, {Rawlings}, {Takahashi}, \&
  {Testi}}]{2013A&A...557A.132M}
{Mathews}, G.~S., {Klaassen}, P.~D., {Juh{\'a}sz}, A., {et~al.} 2013, \aap,
  557, A132

\bibitem[{{Meynet} {et~al.}(2011){Meynet}, {Eggenberger}, \&
  {Maeder}}]{2011A&A...525L..11M}
{Meynet}, G., {Eggenberger}, P., \& {Maeder}, A. 2011, \aap, 525, L11

\bibitem[{{Moran} {et~al.}(1983){Moran}, {Garay}, {Reid}, {Genzel}, {Wright},
  \& {Plambeck}}]{1983ApJ...271L..31M}
{Moran}, J.~M., {Garay}, G., {Reid}, M.~J., {et~al.} 1983, \apjl, 271, L31

\bibitem[{{Moscadelli} {et~al.}(2013){Moscadelli}, {Cesaroni},
  {S{\'a}nchez-Monge}, {Goddi}, {Furuya}, {Sanna}, \&
  {Pestalozzi}}]{2013A&A...558A.145M}
{Moscadelli}, L., {Cesaroni}, R., {S{\'a}nchez-Monge}, {\'A}., {et~al.} 2013,
  \aap, 558, A145

\bibitem[{{Muzerolle} {et~al.}(2013){Muzerolle}, {Furlan}, {Flaherty}, {Balog},
  \& {Gutermuth}}]{2013Natur.493..378M}
{Muzerolle}, J., {Furlan}, E., {Flaherty}, K., {Balog}, Z., \& {Gutermuth}, R.
  2013, \nat, 493, 378

\bibitem[{{Naz{\'e}}(2014)}]{2014psce.conf..340N}
{Naz{\'e}}, Y. 2014, in Putting A Stars into Context: Evolution, Environment,
  and Related Stars, ed. G.~{Mathys}, E.~R. {Griffin}, O.~{Kochukhov},
  R.~{Monier}, \& G.~M. {Wahlgren}, 340--349

\bibitem[{{Peretto} {et~al.}(2013){Peretto}, {Fuller}, {Duarte-Cabral},
  {Avison}, {Hennebelle}, {Pineda}, {Andr{\'e}}, {Bontemps}, {Motte},
  {Schneider}, \& {Molinari}}]{2013A&A...555A.112P}
{Peretto}, N., {Fuller}, G.~A., {Duarte-Cabral}, A., {et~al.} 2013, \aap, 555,
  A112

\bibitem[{{P{\'e}rez} {et~al.}(2014){P{\'e}rez}, {Isella}, {Carpenter}, \&
  {Chandler}}]{2014ApJ...783L..13P}
{P{\'e}rez}, L.~M., {Isella}, A., {Carpenter}, J.~M., \& {Chandler}, C.~J.
  2014, \apjl, 783, L13

\bibitem[{{Pineda} {et~al.}(2012){Pineda}, {Maury}, {Fuller}, {Testi},
  {Garc{\'{\i}}a-Appadoo}, {Peck}, {Villard}, {Corder}, {van Kempen}, {Turner},
  {Tachihara}, \& {Dent}}]{2012A&A...544L...7P}
{Pineda}, J.~E., {Maury}, A.~J., {Fuller}, G.~A., {et~al.} 2012, \aap, 544, L7

\bibitem[{{Qi} {et~al.}(2013){Qi}, {{\"O}berg}, {Wilner}, {D'Alessio},
  {Bergin}, {Andrews}, {Blake}, {Hogerheijde}, \& {van
  Dishoeck}}]{2013Sci...341..630Q}
{Qi}, C., {{\"O}berg}, K.~I., {Wilner}, D.~J., {et~al.} 2013, Science, 341, 630

\bibitem[{{Rodr{\'{\i}}guez} {et~al.}(2012){Rodr{\'{\i}}guez}, {Gonz{\'a}lez},
  {Montes}, {Asiri}, {Raga}, \& {Cant{\'o}}}]{2012ApJ...755..152R}
{Rodr{\'{\i}}guez}, L.~F., {Gonz{\'a}lez}, R.~F., {Montes}, G., {et~al.} 2012,
  \apj, 755, 152

\bibitem[{{Rodr{\'{\i}}guez} {et~al.}(2014){Rodr{\'{\i}}guez}, {Masqu{\'e}},
  {Dzib}, {Loinard}, \& {Kurtz}}]{2014RMxAA..50....3R}
{Rodr{\'{\i}}guez}, L.~F., {Masqu{\'e}}, J.~M., {Dzib}, S.~A., {Loinard}, L.,
  \& {Kurtz}, S.~E. 2014, \rmxaa, 50, 3

\bibitem[{{Rosen} {et~al.}(2012){Rosen}, {Krumholz}, \&
  {Ramirez-Ruiz}}]{2012ApJ...748...97R}
{Rosen}, A.~L., {Krumholz}, M.~R., \& {Ramirez-Ruiz}, E. 2012, \apj, 748, 97

\bibitem[{{Salter} {et~al.}(2008){Salter}, {Hogerheijde}, \&
  {Blake}}]{2008A&A...492L..21S}
{Salter}, D.~M., {Hogerheijde}, M.~R., \& {Blake}, G.~A. 2008, \aap, 492, L21

\bibitem[{{Stauffer} {et~al.}(2014){Stauffer}, {Cody}, {Baglin}, {Alencar},
  {Rebull}, {Hillenbrand}, {Venuti}, {Turner}, {Carpenter}, {Plavchan},
  {Findeisen}, {Carey}, {Terebey}, {Morales-Calder{\'o}n}, {Bouvier}, {Micela},
  {Flaccomio}, {Song}, {Gutermuth}, {Hartmann}, {Calvet}, {Whitney}, {Barrado},
  {Vrba}, {Covey}, {Herbst}, {Furesz}, {Aigrain}, \&
  {Favata}}]{2014AJ....147...83S}
{Stauffer}, J., {Cody}, A.~M., {Baglin}, A., {et~al.} 2014, \aj, 147, 83

\bibitem[{{Szymczak} {et~al.}(2014){Szymczak}, {Wolak}, \&
  {Bartkiewicz}}]{2014MNRAS.439..407S}
{Szymczak}, M., {Wolak}, P., \& {Bartkiewicz}, A. 2014, \mnras, 439, 407

\bibitem[{{van der Marel} {et~al.}(2014){van der Marel}, {van Dishoeck},
  {Bruderer}, \& {van Kempen}}]{2014A&A...563A.113V}
{van der Marel}, N., {van Dishoeck}, E.~F., {Bruderer}, S., \& {van Kempen},
  T.~A. 2014, \aap, 563, A113

\bibitem[{{van der Marel} {et~al.}(2013){van der Marel}, {van Dishoeck},
  {Bruderer}, {Birnstiel}, {Pinilla}, {Dullemond}, {van Kempen}, {Schmalzl},
  {Brown}, {Herczeg}, {Mathews}, \& {Geers}}]{2013Sci...340.1199V}
{van der Marel}, N., {van Dishoeck}, E.~F., {Bruderer}, S., {et~al.} 2013,
  Science, 340, 1199

\bibitem[{{Walsh} {et~al.}(2006){Walsh}, {Bourke}, \&
  {Myers}}]{2006ApJ...637..860W}
{Walsh}, A.~J., {Bourke}, T.~L., \& {Myers}, P.~C. 2006, \apj, 637, 860

\bibitem[{{Wolff} {et~al.}(2006){Wolff}, {Strom}, {Dror}, {Lanz}, \&
  {Venn}}]{2006AJ....132..749W}
{Wolff}, S.~C., {Strom}, S.~E., {Dror}, D., {Lanz}, L., \& {Venn}, K. 2006,
  \aj, 132, 749

\bibitem[{{Zapata} {et~al.}(2013){Zapata}, {Loinard}, {Rodr{\'{\i}}guez},
  {Hern{\'a}ndez-Hern{\'a}ndez}, {Takahashi}, {Trejo}, \&
  {Parise}}]{2013ApJ...764L..14Z}
{Zapata}, L.~A., {Loinard}, L., {Rodr{\'{\i}}guez}, L.~F., {et~al.} 2013,
  \apjl, 764, L14

\end{thebibliography}
\bibliographystyle{apj}

\end{document}